\documentclass[12pt,preprint]{aastex}
\begin{document}

\title{KIC8462852 Faded at an Average Rate of 0.164$\pm$0.013 Magnitudes Per Century From 1890 To 1989}
\author{Bradley E. Schaefer\affil{Physics and Astronomy, Louisiana State University, Baton Rouge, LA 70803}}

\begin{abstract}

KIC8462852 is a completely-ordinary F3 main sequence star, except that the light curve from {\it Kepler} shows episodes of unique and inexplicable day-long dips with up to 20\% dimming.  Here, I provide a light curve of 1338 Johnson B-band magnitudes from 1890 to 1989 taken from archival photographic plates at Harvard.  KIC8462852 displays a secular dimming at an average rate of 0.164$\pm$0.013 magnitudes per century.  From the early-1890s to the late-1980s, KIC8462852 faded by 0.193$\pm$0.030 mag.  The decline is not an artifact because nearby check stars have closely flat light curves.  This century-long dimming is unprecedented for any F-type main sequence star.  Thus the Harvard light curve provides the first confirmation (past the several dips seen in the {\it Kepler} light curve alone) that KIC8462852 has anything unusual.  The century-long dimming and the day-long dips are both just extreme ends of a spectrum of timescales for unique dimming events.  By Ockham's Razor, two such unique and similar effects are very likely produced by one physical mechanism.  This one mechanism does not appear as any isolated catastrophic event in the last century, but rather must be some ongoing process with continuous effects.  Within the context of dust-occultation models, the century-long dimming trend requires $10^4$-$10^7$ times as much dust as for the deepest {\it Kepler} dip.  Within the context of the comet-family idea, the century-long dimming trend requires an estimated 648,000 giant comets (each with 200 km diameter) all orchestrated to pass in front of the star within the last century.

\end{abstract}

\section{Background}

	The star KIC8462852 (TYC 3162-665-1) is apparently a perfectly normal star, with no spectral peculiarities, appearing in the original Cygnus/Lyra field studied with the {\it Kepler} spacecraft.  But then, the {\it Planet Hunters} project discovered in the {\it Kepler} light curve that KIC8462852 displays a unique series of aperiodic dips in brightness (Boyajian et al. 2016).
	
	Boyajian et al. (2016) report a complete study of the properties of the star.  KIC8462852 is a V=11.705 star (B-V=0.557) at 454 parsec distance.  The surface temperature is 6750 K for  spectral type of F3 V with no emission lines or anything unusual.  Critically, the star does not does not have any infrared excess, with this being confirmed by Lisse et al. (2016), Marengo et al. (2015), and Thompson et al. (2016).  With the exception of the {\it DASCH} light curve, all data for KIC8462852 are from after the launch of {\it Kepler} in 2009.  The {\it Kepler} light curve displays a series of dips, where the star faded by 0.2\%--20\% with durations from a day to weeks.
	
	The F3 star appears normal, and such fast variations of a single main sequence star are inexplicable, so attention has been concentrated on the primary star being dimmed by occultation of circumstellar dust clumps.  Boyajian et al. (2016) consider scenarios where the dust originated in a catastrophic collision in an asteroid belt, a giant impact between planets, and a family of comets.  Most of the proposed scenarios are ruled out due to the lack of any infrared excess.  Bodman \& Quillen (2016) investigate the idea of a comet family, but find that they need implausibly-large comets in large numbers, plus a contrived disruption history.  Further, the comet hypothesis cannot explain many of the dip light curves.
	
\section{Photometry With The Harvard Archival Plates}

	The collection of $\approx$500,000 sky photographs in the archives at Harvard College Observatory covers the entire sky from 1890 to 1989.  However, few plates cover the sky from 1953 to 1969 due to the `Menzel Gap'.  A typical glass plate has dimensions of 8$\times$10 inch, stored in a paper envelope on shelves, with angular sizes from 11$\degr$--42$\degr$ wide.  The limiting magnitude varies substantially from plate to plate, with a typical range from B=14 to fainter than B=18.  Any given position is covered by 1000--4000 plates.
	
	The Harvard plates are the original basis for what later became the Johnson B magnitude system.  Despite the changes in emulsions over the years, the color sensitivity of the blue plates has been repeatedly measured to have a negligibly-small color term to the Johnson B system.  For both sets of measures in this paper, the comparison sequence was taken from the AAVSO Photometric All-Sky Survey ({\it APASS}, see Henden \& Munari 2014).  The {\it APASS} magnitudes are accurately tied to the Johnson B system (Munari et al. 2014) through the standard stars of Landolt (2009).
	
	Figure 1 shows two typical plates centered on KIC8462852, with the chosen plates being the first and last, from 1890 and 1989.
	
\subsection{{\it DASCH}}
	
	Many wonderful treasures are saved in the Harvard plates, but the reality is that the current generation of astronomers are mostly unaware of their existence.  J. Grindlay has started and lead the work to completely digitize all $\approx$500,000 plates (Grindlay et al. 2012; Tang et al. 2013).  His program is called {\it Digital Access to a Sky Century @ Harvard} ({\it DASCH})\footnote{http://dasch.rc.fas.harvard.edu/index.php}.  The products are top-quality digitization for each plate, plus fully-calibrated magnitudes for each stellar image on the plate.  Currently, {\it DASCH} has completed only $\approx$15\% of the Harvard archives, including all the plates covering the original Cygnus/Lyra {\it Kepler} field.
	
	Boyajian et al. (2016) extracted the {\it DASCH} light curve for KIC8462852, as part of their collection of data from a wide variety of sources.  They discussed this light curve in four sentences, concluding that ``the star did not do anything spectacular over the past 100 years".  They also concluded that dips as seen with {\it Kepler} would have a high chance of not being visible in the historical Harvard light curve.
	
	The {\it DASCH} analysis pipeline produces either magnitudes or limits for all 1885 plates covering the area of KIC8462852.  For ordinary data quality selection, I reject plates with  (1) `yellow' or `red' sensitive emulsion, (2) quality flags indictor, `AFLAGS', $>$9000, (3) one-sigma error bars $>$0.33 mag, or (4) the target within 0.2 magnitudes of the quoted plate limit.  The AFLAGS value for each image is an array of binary flags, where all the critical quality indicators (e.g., plate defects, near the plate edge, and trailed/overlapped images) are $>$9000.  Along with my other quality cuts, all the plates with AFLAGS$<$9000 are fine under visual magnified inspection.  With these selections, I have 1338 magnitudes from {\it DASCH}.

\subsection{Check Stars}
	
	The traditional and best method to measure the total errors in a light curve is to use identical procedures to construct light curves for check stars.  So I have constructed DASCH light curves with the same quality cuts for 12 check stars.  These stars were selected for (1) the {\it APASS} B-magnitude being within two-thirds of a magnitude of KIC8462852, (2) the {\it APASS} B-V color being within 0.20 mag of KIC8462852, (3) the star being near KIC8462852, with a median distance of 22', and (4) not being crowded with nearby or bright stars.  This last selection is to avoid the known case where some nearby or bright star will bias the photometry on plates with a larger point spread function (PSF).  KIC8462852 has no crowding.  These check stars have identical systematic-plus-measurement errors as KIC8462852.
	
	I had originally selected five nearby check stars. But four of these are greatly different in color or (in one case) much fainter in magnitude than KIC 8462852.  These stars all have the same small RMS scatter of slopes.  (The 5-year-bin slopes were +0.027, -0.018, -0.053, and +0.020 mag/century.)  But there is great strength of using only stars with similar magnitude and similar color as KIC8462852.  With such stars, we have no question or deviation from knowing that all systematic and random effects will be identical for the check stars as for KIC8462852.  With this, I made a sample of 12 check stars as in the previous paragraph that does not include the four stars with the stated slopes.  To fill out the sample to 12, I simply added in the first nearest and uncrowded similar stars that I could identify from the APASS data.
	
	The DASCH light curves for all 12 check stars are given in Table 1.  The average light curve is in Figure 1.  These light curves were fit to a straight line by a chi-square fit, with the given slopes in units of magnitudes-per-century, both for no binning and for 5-year time bins.  The difference in average magnitudes between the decades of the 1890s and the 1980s are also presented as another measure of secular trends.  Over all 12 check stars, the average slope for the 5-year binned light curve is -0.009 with an RMS scatter of 0.043.  For the slopes with no binning, the average is -0.001 with an RMS scatter of 0.040.  The differences in B magnitudes from the decade of the 1890s to the 1980s has an average of -0.007 mag, with an RMS scatter of 0.044 mag.
	
	For evaluating the confidence level for the slope of KIC8462852, we need to check two points.  First, with all the check stars spanning over 1.1 mag in B-V, there is no apparent color term.  And this agrees with the very deep experience of prior Harvard astronomers and myself in looking at light curves of stars of extreme colors on the Harvard plates.  Second, out of the 16 check stars of all colors, 11 are within 1-sigma of the mean (69\%), while the maximum deviations are -1.57 and +1.60 sigma.  With this, the distribution of check star slopes appears to be a good Gaussian distribution, albeit with only 16 stars.
		
	Hippke et al. (2016) have examined DASCH light curves for 41 field stars, stated to all be within 3$\degr$ of KIC8462852.  They claim a substantial scatter in linear slopes.  They report only two specific star identifications and light curves.  For these, their KIC6366512 is 7.0$\degr$ distant, while their KIC9909362 is 3.3$\degr$ distant.  So something is wrong with their selection criterion.  Further, their only sample light curve with a high linear slope (for KIC6366512) is closely crowded with a star of comparable magnitude.  Frequently, on plates with poor PSF, the two star images overlap, and this will necessarily bias the derived magnitudes, so apparent trends in the light curve will simply be produced by changes in the frequency of large PSFs.  For example, on the DNB plate series (all the plates after 1969), KIC6366512 has touching overlap of the saturated regions in 85\% of the plates, so in such cases we would expect a discontinuity across the Menzel Gap.  With their only extreme example being bad, it is easy to expect that their reported large scatter is merely due to bad selections of check stars.

\subsection{Visual Estimates}
	
	Visual estimates are made with the plate placed on a light table, and the star field is examined with magnification provided by a low-power microscope, or more commonly with a loupe placed onto the glass side of the plate.  The image diameter of the target is judged in comparison to the image diameters of each of the comparison stars in turn, and the brightness of the target is evaluated as being some fraction between two comparison stars.  The human eye is remarkably good at such side-by-side comparison of the diameter of small circles.  Inexperienced workers have an accuracy of $\sim$0.3 mag, while experienced observers get to $\sim$0.1 mag accuracy for typical plates and sequences.  I have had very extensive practice at measuring magnitudes at Harvard and plate archives worldwide, continuously from 1979 to the present (e.g., Schaefer 1990; 2014a; 2014b; Schaefer et al. 2008; Schaefer \& Patterson 1983) plus substantial work on the theory of photographic magnitudes (e.g., Schaefer 1979; 1981; 1983; 1995).

	Visual measures have the advantages of being fast, cheap, and simple, whereas scanning methods are always slow, expensive, and complex.  The {\it DASCH} photometry has the advantages that all useable plates were measured, that realistic error bars are calculated for each plate individually, that the photometry is purely objective with no `personal equation', and that nearby check stars can be handled in a manner identical to the target star.  
	
	For the by-eye light curve of KIC8462852, I visited Harvard in October 2015.  I selected plates for a wide distribution in time from the patrol series (DNB, RH, and AC) as well as deep series (A, MC, and I).  For the plates pulled from the shelves and put on a light table, with examination under a 12$\times$ loupe, I continued only for those plates which I judged to be able to return a confident and accurate magnitude.  With this, I measured 131 magnitudes of KIC8462852 from 1890--1989.

\section{KIC8462852 From 1890 to 1989}

	With these procedures, I have created two independent Johnson B light curves for KIC8462852 from 1890 to 1989 from the same set of Harvard plates, the first with 1338 plates with {\it DASCH} magnitudes, and the second with 131 plates with my by-eye measures.  With this, a simple plot of the light curve shows scatter such that it is difficult to pick out dips, eruptions, secular trends, or any variability with amplitude smaller than a quarter of a magnitude or so.  Hence, for display, I have grouped the magnitudes into five year bins.  For the magnitudes in each bin, the averaged magnitude is from a straight average.  For each bin, I have calculated the RMS scatter and I take the one-sigma error bar on the bin average to be the RMS divided by the square root of the number of plates in that bin.  Table 2 and Figure 2 present this KIC8462852 light curve from {\it DASCH}.
	
	With three methods, I find that the average measurement error for individual plates is close to 0.13 mag:  (1) For the {\it DASCH} light curve, in the 5-year time bins, the average RMS scatter of the individual magnitudes is 0.12 mag.  (2) {\it DASCH} calculates realistic error bars for individual plates.  For KIC8462852, the median is 0.15 mag, with a central 68\% range of 0.10-0.21 mag.  (3) The differences between by-eye and {\it DASCH} magnitudes of the same plate have an RMS of 0.19 mag, so the average one-sigma uncertainty in one measure of the plate is $0.19/\sqrt{2}$=0.13 mag.
	
	The DASCH light curve displays highly significant variations, with a clear trend for fading from early to late times.  A chi-square fit for a linear trend in the 5-year binned light curve has a slope of +0.164$\pm$0.013 magnitudes per century.  With no time binning, the slope is +0.150$\pm$0.014 mag/century.  From the first time bin to the last, the light curve has faded 0.193$\pm$0.030 mag.  For the decades of the 1890s and the 1980s, the change in brightness is 0.183$\pm$0.016 mag. 
	
	My by-eye light curve also has an obvious slope.  A chi-square fit to all 131 magnitudes from 1890 to 1989 yields a slope of +0.324$\pm$0.040 magnitudes per century.  This is formally different from the slope that I get from {\it DASCH}, and I attribute this to the happenstance that my sampling of the available plates included few from 1900-1909, when the light curve was relatively dim and pulling the fitted slope to smaller values.  The critical point from my by-eye measures is that KIC8462852 does indeed have a highly significant variation, manifesting as a secular fading from the 1890s to the 1980s.  This proves that the secular trend is not due to any issues with the {\it DASCH} procedures, measures, analysis, or selection.
	
	The significance of the linear slope in the 5-year binned light curve is (0.164+0.009)/0.043, or 4.0-sigma.  For no time binning, the significance is 3.8-sigma.  For the difference from the 1890s to the 1980s, the significance is 4.0-sigma.  For the visual light curve, the significance is from 8-sigma to 12-sigma for the three measures.
	
	The long-term trend in the {\it DASCH} light curve can be described in various ways.  One way is simply to note that KIC8462852 faded from B=12.265$\pm$0.028 in 1892.5 to B=12.458$\pm$0.012 in 1987.5, for a total fading of 0.193$\pm$0.030 mag in 95 years (+0.203$\pm$0.032 mag/century).  This end-to-end trend line provides an excellent representation of all the Harvard data {\it except} for the decade from 1900-1909.  This might be due to the star suffering many deep dips during the years 1900-1909.  I am not claiming that the secular decline is steady or monotonic, and the 1900-1909 minimum is not of high significance.  The light curve has an alternative description that it has a secular trend that is not steady.  That is, the average decline rate is +0.164$\pm$0.013 magnitudes per century, but there are stutters built on top of this.
	
\section{Implications}

	The KIC8462852 light curve from 1890 to 1989 shows a highly significant secular trend in fading over 100 years, with this being completely unprecedented for any F-type main sequence star.  Such stars should be very stable in brightness, with evolution making for changes only on time scales of many millions of years.  So the Harvard data alone prove that KIC8462852 has unique and large-amplitude photometric variations.  Previously, the {\it only} evidence that KIC8462852 was unusual in any way was a few dips in magnitude as observed by one satellite, so inevitably we have to wonder whether the whole story is just some problem with {\it Kepler}.  Boyajian et al. (2016) had already made a convincing case that the dips were not caused by any data or analysis artifacts, and their case is strong.  Nevertheless, it is comforting to know from two independent sources that KIC8462852 is displaying unique and inexplicable photometric variations.
	
	KIC8462852 is suffering a century-long secular fading, and this is contrary to the the various speculation that the obscuring dust was created by some singular catastrophic event.  If any such singular event happened after around 1920, then the prior light curve should appear perfectly flat, whereas there is significant variability before 1920.  If the trend is caused by multiple small catastrophic events, then it is difficult to understand how they can time themselves so as to mimic the trend from 1890-1989.  In the context of the idea that the star in undergoing a Late Heavy Bombardment (Lisse et al. 2015), it is implausible that such a mechanism could start up on a time scale of a century, or that it would start so smoothly with many well-spaced collisions.
	
	KIC8462852 displays {\it two} types of unique dimming episodes that differ only in the time scale (the dips from {\it Kepler} and the fading from Harvard) and these must be causally related and coming from the same mechanism.  That is, Ockham's Razor tells us that it is very unlikely that one star will suffer two different mechanisms that are unique to that star and that both are only manifest in dimming the starlight by up to 20\%.  The timescales differ greatly, from a day for the {\it Kepler} dips up to a century for the Harvard light curve fading.  However, dimming events with intermediate timescales are also seen (e.g., the 1900-1909 decade and the last hundred days of the {\it Kepler} light curve), so apparently there is a continuum of timescales available for the one dimming mechanism.  That is, the single mechanism must have some feature that can vary so as to produce the continuum of timescales from a day to a century.  So if the day-long dips are caused by circumstellar dust occultations, then the century-long fading are also caused by circumstellar dust occultations.  
	
	Within the various dust-occultation ideas, there is some quantity of dust ($M_{dust,1dip}$) required to create the one deepest dip (20\% extinction with a duration of around one day), which only dims the star from the {\it Kepler} baseline level.   Boyajian et al. (2016) and Thompson et al. (2016) calculate a lower limit on $M_{dust,1dip}$ of $10^{-9}$ $M_{\oplus}$.  Bodman \& Quillen (2016) calculate that the comet family scenario requires 36 giant-comets with 200 km diameters to produce enough dust.  
	
	The no-circumstellar-extinction level of the star is $>$0.193$\pm$0.030 mag brighter in the 1890s than is the {\it Kepler} baseline, so any dust-occultation idea also requires that the star be covered by a second portion of dust, roughly $M_{dust,1dip}$, just to dim the star by around 20\% down to the {\it Kepler} baseline.  The time for this given mass of dust to cross over the star is of order one day (as based on the dip duration), so in the day after the deepest {\it Kepler} dip there must be some fresh supply of dust that keeps the star dimmed to the {\it Kepler} baseline.  Over the whole 1500 days of the {\it Kepler} light curve, the total dust needed will be 1503 $M_{dust,1dip}$ with 1500 $M_{dust,1dip}$ simply to dim the star down to the {\it Kepler} baseline plus 3 $M_{dust,1dip}$ to produce all the {\it Kepler} dips.  The dust extinction from 1890 observed in the Harvard plates is not constant, and gets to near 20\% below the brightest level only towards the end of the century.  With a linear decline, the required dust production would be equivalent to the full 20\% for half a century.  This will provide a lower limit, since it has ignored the extra dimming from 1900-1909, while the real no-circumstellar-extinction level of the star could well be brighter than seen in the 1890s.  Half a century is roughly 18,000 days, so the fading as seen with the Harvard plates requires $>$18,000 $M_{dust,1dip}$.  And that is just for the dust that has happened to pass in front of the star.  If the star's dust inventory is not confined to orbits that have it all passing in front of the star in the last century, then some roughly spherical distribution can be expected.  For dust at a distance of $R$ from the star, the total dust mass would then be $(R/R_*)^2 \times M_{dust,1dip}$ just to dim the star down to the {\it Kepler} baseline.  The radius of the star ($R_*$) is 1.58 $R_{\odot}$ and the dust is from 3--30 AU from the star (Boyajian et al. 2016), so some roughly isotropic dust distribution will require 170,000 to 17,000,000 times $M_{dust,1dip}$.  In all, the dimming shown in the Harvard light curve requires that there be of order $10^4$ to $10^7$ times as much dust as has been previously modeled from the {\it Kepler} dips alone.	
	
	With $M_{dust,1dip}\gtrsim10^{-9}$ $M_{\oplus}$, the dimming of the Harvard light curve requires of order $10^{-5}$ to $10^{-2}$ $M_{\oplus}$ of dust around KIC8462852.  Thompson et al. (2016) have used SCUBA-2 sub-millimeter observations to place limits on the total dust mass around the star, with a limit of $\leq$3.0$\times$$10^{-6}$ $M_{\oplus}$ for dust 2--8 AU from the star and a limit of $\leq$5.6$\times$$10^{-3}$ $M_{\oplus}$ for dust out to 26 AU from the star.  The only way to reconcile these limits with the fading in the Harvard light curve, is to require that the dust be confined to a volume around a plane (like for an orbit or a disk) and/or to be far from the star.

	With 36 giant-comets required to make the one 20\% {\it Kepler} dip, and all of these along one orbit, we would need 648,000 giant-comets to create the century-long fading.  For these 200 km diameter giant-comets having a density of 1 gm cm$^{-3}$, each will have a mass of $4 \times 10^{21}$ gm, and the total will have a mass of 0.4 $M_{\oplus}$.  This can be compared to the largest known comet in our own Solar System (Comet Hale-Bopp) with a diameter of 60 km.  This can also be compared to the entire mass of the Kuiper Belt at around 0.1 $M_{\oplus}$ (Gladman et al. 2001).  I do not see how it is possible for something like 648,000 giant-comets to exist around one star, nor to have their orbits orchestrated so as to all pass in front of the star within the last century.  So I take this century-long dimming as a strong argument against the comet-family hypothesis to explain the {\it Kepler} dips.
	
	~
	
	The DASCH project has support from NSF grants AST-0407380, AST-0909073, and AST-1313370.  I thank J. Grindlay for detailed checks of the DASCH light curves, and T. Boyajian for detailed checks on the manuscript and the science.

{}

\begin{deluxetable}{lllll}
\tabletypesize{\scriptsize}
\tablecaption{DASCH light curves of twelve check stars
\label{tbl2}}
\tablewidth{0pt}
\tablehead{
\colhead{Year}   &
\colhead{$\langle$B$\rangle$ ($N_{obs}$)}   &
\colhead{$\langle$B$\rangle$ ($N_{obs}$)}   &
\colhead{$\langle$B$\rangle$ ($N_{obs}$)}   &
\colhead{$\langle$B$\rangle$ ($N_{obs}$)}
}
\startdata

			&	\textbf{	TYC 3162-1001-1					:}	&	\textbf{	TYC 3162-1679-1					:}	&	\textbf{	TYC 3162-1320-1					:}	&	\textbf{	TYC 3162-509-1					:}	\\
1892.5	$\pm$	2.5	&		11.713	$\pm$	0.025	(	15	)	&		12.056	$\pm$	0.019	(	15	)	&		12.117	$\pm$	0.032	(	15	)	&		12.196	$\pm$	0.037	(	15	)	\\
1897.5	$\pm$	2.5	&		11.752	$\pm$	0.016	(	49	)	&		12.054	$\pm$	0.016	(	31	)	&		12.112	$\pm$	0.014	(	37	)	&		12.145	$\pm$	0.016	(	46	)	\\
1902.5	$\pm$	2.5	&		11.768	$\pm$	0.010	(	148	)	&		12.034	$\pm$	0.014	(	122	)	&		12.160	$\pm$	0.012	(	114	)	&		12.184	$\pm$	0.015	(	110	)	\\
1907.5	$\pm$	2.5	&		11.825	$\pm$	0.019	(	62	)	&		12.013	$\pm$	0.024	(	46	)	&		12.140	$\pm$	0.029	(	35	)	&		12.225	$\pm$	0.052	(	27	)	\\
1912.5	$\pm$	2.5	&		11.770	$\pm$	0.009	(	117	)	&		12.048	$\pm$	0.013	(	110	)	&		12.158	$\pm$	0.013	(	100	)	&		12.186	$\pm$	0.015	(	95	)	\\
1917.5	$\pm$	2.5	&		11.781	$\pm$	0.014	(	88	)	&		12.073	$\pm$	0.016	(	89	)	&		12.170	$\pm$	0.017	(	67	)	&		12.165	$\pm$	0.017	(	81	)	\\
1922.5	$\pm$	2.5	&		11.773	$\pm$	0.013	(	51	)	&		12.073	$\pm$	0.015	(	57	)	&		12.163	$\pm$	0.030	(	51	)	&		12.197	$\pm$	0.023	(	44	)	\\
1927.5	$\pm$	2.5	&		11.791	$\pm$	0.012	(	118	)	&		12.059	$\pm$	0.015	(	119	)	&		12.141	$\pm$	0.015	(	105	)	&		12.219	$\pm$	0.018	(	92	)	\\
1932.5	$\pm$	2.5	&		11.776	$\pm$	0.014	(	108	)	&		12.044	$\pm$	0.014	(	125	)	&		12.153	$\pm$	0.017	(	105	)	&		12.220	$\pm$	0.014	(	97	)	\\
1937.5	$\pm$	2.5	&		11.758	$\pm$	0.008	(	241	)	&		12.050	$\pm$	0.008	(	255	)	&		12.154	$\pm$	0.008	(	236	)	&		12.216	$\pm$	0.010	(	228	)	\\
1942.5	$\pm$	2.5	&		11.759	$\pm$	0.009	(	190	)	&		12.048	$\pm$	0.008	(	228	)	&		12.164	$\pm$	0.010	(	211	)	&		12.201	$\pm$	0.009	(	197	)	\\
1947.5	$\pm$	2.5	&		11.772	$\pm$	0.013	(	97	)	&		12.041	$\pm$	0.009	(	167	)	&		12.129	$\pm$	0.010	(	156	)	&		12.169	$\pm$	0.011	(	145	)	\\
1952.5	$\pm$	2.5	&		11.760	$\pm$	0.014	(	66	)	&		12.045	$\pm$	0.019	(	69	)	&		12.162	$\pm$	0.023	(	66	)	&		12.208	$\pm$	0.018	(	66	)	\\
1962.5	$\pm$	2.5	&		11.755	$\pm$	0.045	(	6	)	&		12.208	$\pm$	0.066	(	5	)	&		12.100	$\pm$	0.060	(	4	)	&		12.164	$\pm$	0.054	(	5	)	\\
1967.5	$\pm$	2.5	&		11.756	$\pm$	0.065	(	5	)	&		11.968	$\pm$	0.065	(	6	)	&		12.138	$\pm$	0.068	(	5	)	&		12.207	$\pm$	0.049	(	6	)	\\
1972.5	$\pm$	2.5	&		11.751	$\pm$	0.019	(	15	)	&		12.084	$\pm$	0.021	(	15	)	&		12.166	$\pm$	0.040	(	16	)	&		12.154	$\pm$	0.037	(	14	)	\\
1977.5	$\pm$	2.5	&		11.761	$\pm$	0.014	(	52	)	&		12.056	$\pm$	0.016	(	48	)	&		12.163	$\pm$	0.015	(	59	)	&		12.118	$\pm$	0.015	(	56	)	\\
1982.5	$\pm$	2.5	&		11.724	$\pm$	0.011	(	83	)	&		12.057	$\pm$	0.010	(	82	)	&		12.174	$\pm$	0.012	(	93	)	&		12.128	$\pm$	0.012	(	81	)	\\
1987.5	$\pm$	2.5	&	\underline{	11.761	$\pm$	0.012	(	100	)}	&	\underline{	12.064	$\pm$	0.011	(	95	)}	&	\underline{	12.184	$\pm$	0.014	(	108	)}	&	\underline{	12.178	$\pm$	0.013	(	96	)}	\\
Number mags =			&		1611						&		1684						&		1583						&		1501						\\
Slope$_{\rm no binning}$ =			&		0.002	$\pm$	0.013				&		-0.006	$\pm$	0.013				&		0.020	$\pm$	0.013				&		0.022	$\pm$	0.013				\\
Slope$_{\rm 5 year bins}$ =			&		-0.029	$\pm$	0.011				&		0.014	$\pm$	0.011				&		0.034	$\pm$	0.013				&		-0.047	$\pm$	0.013				\\
$\langle$B$\rangle_{1890s}$-$\langle$B$\rangle_{1980s}$ =			&		-0.002	$\pm$	0.016				&		-0.005	$\pm$	0.014				&		-0.066	$\pm$	0.016				&		0.002	$\pm$	0.017				\\
			&								&								&								&								\\
			&	\textbf{	TYC 3559-1478-1					:}	&	\textbf{	J200626.1+440816					:}	&	\textbf{	TYC 3162-857-1					:}	&	\textbf{	TYC 3162-821-1					:}	\\
1892.5	$\pm$	2.5	&		12.074	$\pm$	0.021	(	18	)	&		12.014	$\pm$	0.053	(	13	)	&		12.172	$\pm$	0.033	(	19	)	&		11.917	$\pm$	0.025	(	21	)	\\
1897.5	$\pm$	2.5	&		12.071	$\pm$	0.011	(	54	)	&		11.959	$\pm$	0.018	(	39	)	&		12.138	$\pm$	0.015	(	42	)	&		11.900	$\pm$	0.020	(	52	)	\\
1902.5	$\pm$	2.5	&		12.067	$\pm$	0.012	(	129	)	&		11.981	$\pm$	0.012	(	126	)	&		12.174	$\pm$	0.013	(	119	)	&		11.925	$\pm$	0.014	(	131	)	\\
1907.5	$\pm$	2.5	&		12.111	$\pm$	0.025	(	51	)	&		11.967	$\pm$	0.023	(	39	)	&		12.211	$\pm$	0.025	(	28	)	&		11.930	$\pm$	0.023	(	50	)	\\
1912.5	$\pm$	2.5	&		12.033	$\pm$	0.014	(	112	)	&		11.993	$\pm$	0.011	(	105	)	&		12.178	$\pm$	0.013	(	106	)	&		11.919	$\pm$	0.013	(	116	)	\\
1917.5	$\pm$	2.5	&		12.038	$\pm$	0.014	(	94	)	&		11.978	$\pm$	0.017	(	88	)	&		12.163	$\pm$	0.015	(	81	)	&		11.921	$\pm$	0.015	(	90	)	\\
1922.5	$\pm$	2.5	&		12.032	$\pm$	0.020	(	57	)	&		11.989	$\pm$	0.015	(	62	)	&		12.191	$\pm$	0.019	(	49	)	&		11.937	$\pm$	0.021	(	59	)	\\
1927.5	$\pm$	2.5	&		12.023	$\pm$	0.014	(	115	)	&		12.044	$\pm$	0.014	(	122	)	&		12.191	$\pm$	0.015	(	102	)	&		11.957	$\pm$	0.014	(	123	)	\\
1932.5	$\pm$	2.5	&		12.007	$\pm$	0.013	(	112	)	&		11.998	$\pm$	0.011	(	125	)	&		12.175	$\pm$	0.016	(	111	)	&		11.963	$\pm$	0.013	(	123	)	\\
1937.5	$\pm$	2.5	&		12.033	$\pm$	0.008	(	261	)	&		11.998	$\pm$	0.008	(	270	)	&		12.173	$\pm$	0.008	(	245	)	&		11.936	$\pm$	0.008	(	265	)	\\
1942.5	$\pm$	2.5	&		12.037	$\pm$	0.008	(	237	)	&		12.000	$\pm$	0.008	(	244	)	&		12.196	$\pm$	0.009	(	193	)	&		11.936	$\pm$	0.009	(	219	)	\\
1947.5	$\pm$	2.5	&		12.029	$\pm$	0.009	(	167	)	&		11.998	$\pm$	0.010	(	178	)	&		12.151	$\pm$	0.011	(	151	)	&		11.917	$\pm$	0.010	(	158	)	\\
1952.5	$\pm$	2.5	&		12.029	$\pm$	0.018	(	70	)	&		12.029	$\pm$	0.017	(	75	)	&		12.178	$\pm$	0.024	(	58	)	&		11.909	$\pm$	0.017	(	74	)	\\
1962.5	$\pm$	2.5	&		12.043	$\pm$	0.060	(	4	)	&		12.013	$\pm$	0.076	(	6	)	&		12.170	$\pm$	0.069	(	3	)	&		11.840	$\pm$	0.060	(	4	)	\\
1967.5	$\pm$	2.5	&		12.036	$\pm$	0.054	(	5	)	&		11.975	$\pm$	0.060	(	4	)	&		12.177	$\pm$	0.086	(	6	)	&		11.895	$\pm$	0.053	(	6	)	\\
1972.5	$\pm$	2.5	&		12.007	$\pm$	0.031	(	16	)	&		11.977	$\pm$	0.036	(	15	)	&		12.154	$\pm$	0.042	(	13	)	&		11.972	$\pm$	0.033	(	13	)	\\
1977.5	$\pm$	2.5	&		12.009	$\pm$	0.013	(	60	)	&		12.010	$\pm$	0.015	(	54	)	&		12.207	$\pm$	0.015	(	58	)	&		11.879	$\pm$	0.014	(	56	)	\\
1982.5	$\pm$	2.5	&		12.002	$\pm$	0.010	(	88	)	&		12.042	$\pm$	0.011	(	81	)	&		12.212	$\pm$	0.012	(	84	)	&		11.854	$\pm$	0.014	(	83	)	\\
1987.5	$\pm$	2.5	&	\underline{	12.001	$\pm$	0.011	(	103	)}	&	\underline{	12.022	$\pm$	0.013	(	97	)}	&	\underline{	12.173	$\pm$	0.013	(	98	)}	&	\underline{	11.914	$\pm$	0.012	(	100	)}	\\
Number mags =			&		1753						&		1743						&		1566						&		1743						\\
Slope$_{\rm no binning}$ =			&		-0.088	$\pm$	0.013				&		0.054	$\pm$	0.013				&		0.037	$\pm$	0.013				&		-0.003	$\pm$	0.013				\\
Slope$_{\rm 5 year bins}$ =			&		-0.070	$\pm$	0.011				&		0.056	$\pm$	0.012				&		0.031	$\pm$	0.012				&		-0.046	$\pm$	0.013				\\
$\langle$B$\rangle_{1890s}$-$\langle$B$\rangle_{1980s}$ =			&		0.070	$\pm$	0.012				&		-0.068	$\pm$	0.019				&		-0.051	$\pm$	0.016				&		0.017	$\pm$	0.018				\\
			&								&								&								&								\\
			&	\textbf{	TYC 3162-488-1					:}	&	\textbf{	TYC 3162-823-1					:}	&	\textbf{	TYC 3162-322-1					:}	&	\textbf{	TYC 3162-1122-1					:}	\\
1892.5	$\pm$	2.5	&		12.162	$\pm$	0.036	(	11	)	&		12.234	$\pm$	0.035	(	16	)	&		$\ldots$			(	0	)	&		12.268	$\pm$	0.023	(	13	)	\\
1897.5	$\pm$	2.5	&		12.160	$\pm$	0.015	(	31	)	&		12.240	$\pm$	0.017	(	44	)	&		$\ldots$			(	0	)	&		12.240	$\pm$	0.016	(	27	)	\\
1902.5	$\pm$	2.5	&		12.145	$\pm$	0.012	(	104	)	&		12.219	$\pm$	0.014	(	103	)	&		12.484	$\pm$	0.051	(	17	)	&		12.251	$\pm$	0.013	(	101	)	\\
1907.5	$\pm$	2.5	&		12.155	$\pm$	0.030	(	36	)	&		12.227	$\pm$	0.024	(	24	)	&		12.244	$\pm$	0.130	(	5	)	&		12.155	$\pm$	0.040	(	24	)	\\
1912.5	$\pm$	2.5	&		12.156	$\pm$	0.014	(	94	)	&		12.257	$\pm$	0.013	(	95	)	&		12.465	$\pm$	0.016	(	77	)	&		12.270	$\pm$	0.015	(	95	)	\\
1917.5	$\pm$	2.5	&		12.149	$\pm$	0.016	(	80	)	&		12.228	$\pm$	0.018	(	77	)	&		12.417	$\pm$	0.016	(	56	)	&		12.275	$\pm$	0.016	(	74	)	\\
1922.5	$\pm$	2.5	&		12.144	$\pm$	0.014	(	53	)	&		12.223	$\pm$	0.018	(	44	)	&		12.408	$\pm$	0.048	(	16	)	&		12.240	$\pm$	0.016	(	37	)	\\
1927.5	$\pm$	2.5	&		12.170	$\pm$	0.015	(	100	)	&		12.225	$\pm$	0.017	(	83	)	&		12.446	$\pm$	0.029	(	46	)	&		12.284	$\pm$	0.016	(	90	)	\\
1932.5	$\pm$	2.5	&		12.144	$\pm$	0.015	(	96	)	&		12.250	$\pm$	0.015	(	101	)	&		12.397	$\pm$	0.019	(	66	)	&		12.258	$\pm$	0.017	(	101	)	\\
1937.5	$\pm$	2.5	&		12.144	$\pm$	0.010	(	228	)	&		12.237	$\pm$	0.010	(	224	)	&		12.462	$\pm$	0.013	(	182	)	&		12.266	$\pm$	0.010	(	227	)	\\
1942.5	$\pm$	2.5	&		12.136	$\pm$	0.010	(	222	)	&		12.240	$\pm$	0.011	(	180	)	&		12.433	$\pm$	0.011	(	144	)	&		12.249	$\pm$	0.008	(	190	)	\\
1947.5	$\pm$	2.5	&		12.116	$\pm$	0.012	(	160	)	&		12.224	$\pm$	0.010	(	144	)	&		12.365	$\pm$	0.015	(	105	)	&		12.251	$\pm$	0.010	(	126	)	\\
1952.5	$\pm$	2.5	&		12.160	$\pm$	0.024	(	53	)	&		12.201	$\pm$	0.017	(	66	)	&		12.447	$\pm$	0.024	(	39	)	&		12.293	$\pm$	0.019	(	45	)	\\
1962.5	$\pm$	2.5	&		12.195	$\pm$	0.064	(	4	)	&		12.220	$\pm$	0.060	(	4	)	&		12.343	$\pm$	0.069	(	3	)	&		12.230	$\pm$	0.050	(	5	)	\\
1967.5	$\pm$	2.5	&		12.200	$\pm$	0.045	(	7	)	&		12.178	$\pm$	0.060	(	4	)	&		12.416	$\pm$	0.054	(	5	)	&		12.325	$\pm$	0.036	(	11	)	\\
1972.5	$\pm$	2.5	&		12.212	$\pm$	0.033	(	13	)	&		12.286	$\pm$	0.031	(	9	)	&		12.270	$\pm$	0.085	(	2	)	&		12.314	$\pm$	0.038	(	15	)	\\
1977.5	$\pm$	2.5	&		12.174	$\pm$	0.017	(	48	)	&		12.245	$\pm$	0.020	(	50	)	&		12.437	$\pm$	0.027	(	16	)	&		12.214	$\pm$	0.017	(	49	)	\\
1982.5	$\pm$	2.5	&		12.198	$\pm$	0.012	(	73	)	&		12.226	$\pm$	0.013	(	82	)	&		12.451	$\pm$	0.043	(	16	)	&		12.165	$\pm$	0.015	(	73	)	\\
1987.5	$\pm$	2.5	&	\underline{	12.175	$\pm$	0.015	(	92	)}	&	\underline{	12.242	$\pm$	0.012	(	101	)}	&	\underline{	12.441	$\pm$	0.026	(	33	)}	&	\underline{	12.231	$\pm$	0.014	(	95	)}	\\
Number mags =			&		1505						&		1451						&		828						&		1398						\\
Slope$_{\rm no binning}$ =			&		0.020	$\pm$	0.014				&		-0.006	$\pm$	0.014				&		-0.003	$\pm$	0.027				&		-0.063	$\pm$	0.014				\\
Slope$_{\rm 5 year bins}$ =			&		0.038	$\pm$	0.013				&		0.001	$\pm$	0.013				&		-0.033	$\pm$	0.027				&		-0.047	$\pm$	0.014				\\
$\langle$B$\rangle_{1890s}$-$\langle$B$\rangle_{1980s}$ =			&		-0.029	$\pm$	0.017				&		0.004	$\pm$	0.018				&		$\ldots$						&		0.048	$\pm$	0.017				\\

\enddata
\end{deluxetable}

\begin{deluxetable}{lll}
\tabletypesize{\scriptsize}
\tablecaption{Harvard light curves of KIC8462852
\label{tbl1}}
\tablewidth{0pt}
\tablehead{
\colhead{}   &
\colhead{KIC8462852}   &
\colhead{KIC8462852}\\
\colhead{}   &
\colhead{DASCH}   &
\colhead{Visual}\\
\colhead{Year}   &
\colhead{$\langle$B$\rangle$ ($N_{obs}$)}   &
\colhead{$\langle$B$\rangle$ ($N_{obs}$)}
}
\startdata

1892.5	$\pm$	2.5	&	12.265	$\pm$	0.028	 (	13	)&	12.177	$\pm$	0.081	 (	15	)	\\
1897.5	$\pm$	2.5	&	12.277	$\pm$	0.015	 (	41	)&	12.190	$\pm$	0.043	 (	21	)	\\
1902.5	$\pm$	2.5	&	12.382	$\pm$	0.017	 (	95	)&	12.450	$\pm$	0.103	 (	5	)	\\
1907.5	$\pm$	2.5	&	12.375	$\pm$	0.040	 (	17	)&	12.160	$\pm$	0.200	 (	1	)	\\
1912.5	$\pm$	2.5	&	12.317	$\pm$	0.014	 (	91	)&	12.290	$\pm$	0.087	 (	13	)	\\
1917.5	$\pm$	2.5	&	12.346	$\pm$	0.016	 (	67	)&	12.386	$\pm$	0.056	 (	17	)	\\
1922.5	$\pm$	2.5	&	12.353	$\pm$	0.017	 (	32	)&	12.303	$\pm$	0.097	 (	3	)	\\
1927.5	$\pm$	2.5	&	12.360	$\pm$	0.019	 (	72	)&	...			 (	0	)	\\
1932.5	$\pm$	2.5	&	12.345	$\pm$	0.022	 (	69	)&	12.800	$\pm$	0.200	 (	1	)	\\
1937.5	$\pm$	2.5	&	12.352	$\pm$	0.008	 (	204	)&	12.400	$\pm$	0.200	 (	1	)	\\
1942.5	$\pm$	2.5	&	12.354	$\pm$	0.009	 (	175	)&	12.545	$\pm$	0.067	 (	2	)	\\
1947.5	$\pm$	2.5	&	12.327	$\pm$	0.010	 (	146	)&	12.300	$\pm$	0.200	 (	1	)	\\
1952.5	$\pm$	2.5	&	12.370	$\pm$	0.020	 (	70	)&	12.346	$\pm$	0.063	 (	12	)	\\
1962.5	$\pm$	2.5	&	12.268	$\pm$	0.175	 (	4	)&	12.343	$\pm$	0.112	 (	3	)	\\
1967.5	$\pm$	2.5	&	12.432	$\pm$	0.030	 (	6	)&	12.442	$\pm$	0.007	 (	5	)	\\
1972.5	$\pm$	2.5	&	12.430	$\pm$	0.040	 (	9	)&	12.560	$\pm$	0.200	 (	1	)	\\
1977.5	$\pm$	2.5	&	12.431	$\pm$	0.018	 (	52	)&	12.435	$\pm$	0.095	 (	2	)	\\
1982.5	$\pm$	2.5	&	12.456	$\pm$	0.012	 (	83	)&	12.598	$\pm$	0.032	 (	12	)	\\
1987.5	$\pm$	2.5	&	12.458	$\pm$	0.012	 (	92	)&	12.561	$\pm$	0.032	 (	16	)	\\
			&						&							\\
Number mags=			&					1338	&					131		\\
Slope$_{no binning}$			&	0.150	$\pm$	0.014			&	0.357	$\pm$	0.030				\\
Slope$_{5 year bins}$			&	0.164	$\pm$	0.013			&	0.324	$\pm$	0.040				\\
$\langle$B$\rangle_{1890s}$-$\langle$B$\rangle_{1980s}$			&	-0.183	$\pm$	0.016			&	-0.393	$\pm$	0.044				\\

\enddata
\end{deluxetable}

\begin{figure}
	\centering
	\makebox[\textwidth][c]{\includegraphics[width=0.7\textwidth]{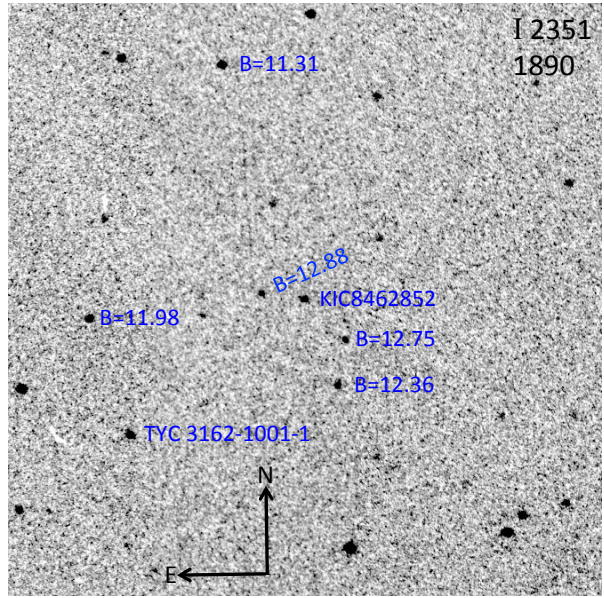}}
	\makebox[\textwidth][c]{\includegraphics[width=0.7\textwidth]{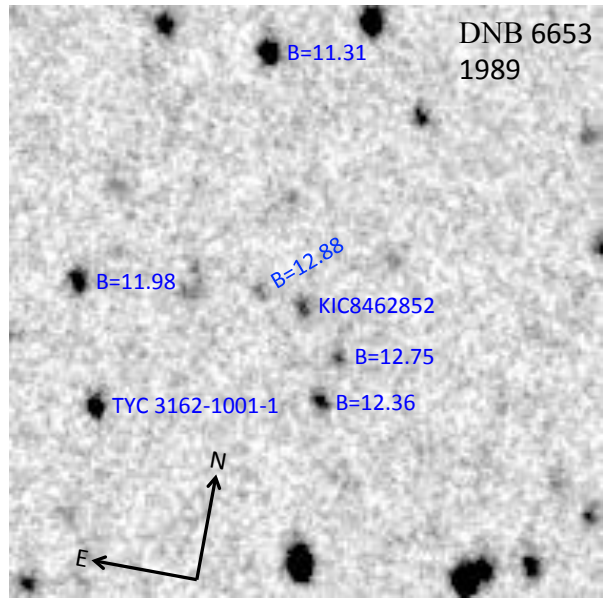}}
	\caption{DASCH scans of the first plate (in 1890) and the last plate (in 1989).  The boxes are 10 arc-minutes on a side, the labeled B magnitudes for comparison stars are from APASS, and only one check star (TYC 3162-1001-1) is in the field.  The first panel for plate I2351 has a reported DASCH magnitude of B=12.24$\pm$0.14, while the second panel for plate DNB6653 has a reported DASCH magnitude of B=12.50$\pm$0.17.}
\end{figure}

\begin{figure}
	\centering
	\makebox[\textwidth][c]{\includegraphics[width=1.\textwidth]{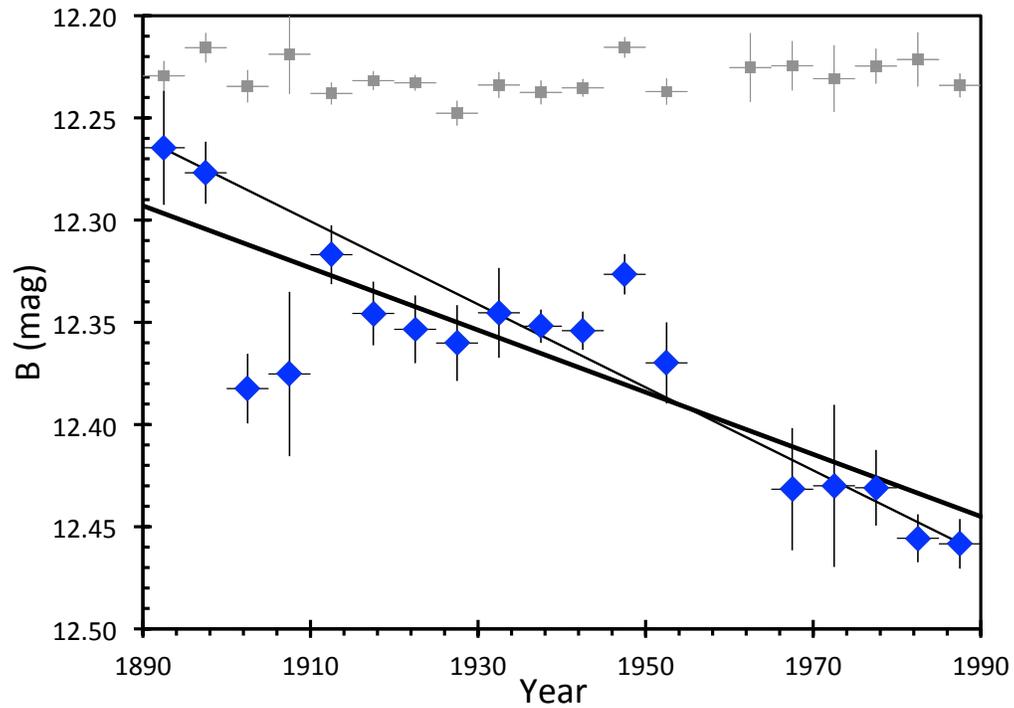}}
	\caption{The 5-year binned {\it DASCH} light curve of KIC8462852 (large blue diamonds).  The star shows highly significant fading from 1890 to 1989.  The averaged light curves for all 12 check stars is displayed, with a small vertical offset, in the figure with grey squares.  The thin line is a simple linear trend connecting the two end points with a slope of 0.203$\pm$0.032 mag/century, while the thick line is the chi-square fit result with a slope of 0.164$\pm$0.013 mag/century.}
\end{figure}

\end{document}